\documentclass[useAMS,usenatbib]{mn2e}

\usepackage{graphics,graphicx}
\usepackage{aas_macros}
\usepackage{amssymb}
\usepackage{latexsym}
\usepackage{color}
\usepackage{epstopdf}
\usepackage{marvosym}

\newcommand{\cm}{cm$^{-1}$}

\begin{document}

\title[Collisional excitation of NH$_3$]{Collisional excitation of
  NH$_3$ by atomic and molecular hydrogen} \author[Bouhafs et
  al.]{N. Bouhafs$^1$, C. Rist$^{2}$, F. Daniel$^{2}$,
  F. Dumouchel$^{1}$, F. Lique$^{1}$, L. Wiesenfeld$^{2}$ \newauthor
  and A. Faure$^{2}$\thanks{E-mail:
    alexandre.faure@univ-grenoble-alpes.fr}\\ $^{1}$Universit\'e
  Normandie, CNRS, LOMC, F-76058 Le Havre, France\\ $^{2}$Universit\'e
  Grenoble Alpes, CNRS, IPAG, F-38000 Grenoble, France }

\pagerange{\pageref{firstpage}--\pageref{lastpage}} \pubyear{2014}

\maketitle

\label{firstpage}

\begin{abstract}

We report extensive theoretical calculations on the rotation-inversion
excitation of interstellar ammonia (NH$_3$) due to collisions with
atomic and molecular hydrogen (both para- and
ortho-H$_2$). Close-coupling calculations are performed for total
energies in the range 1-2000~cm$^{-1}$ and rotational cross sections
are obtained for all transitions among the lowest 17 and 34
rotation-inversion levels of ortho- and para-NH$_3$,
respectively. Rate coefficients are deduced for kinetic temperatures
up to 200~K. Propensity rules for the three colliding partners are
discussed and we also compare the new results to previous calculations
for the spherically symmetrical He and para-H$_2$ projectiles.
Significant differences are found between the different sets of
calculations. Finally, we test the impact of the new rate coefficients
on the calibration of the ammonia thermometer. We find that the
calibration curve is only weakly sensitive to the colliding partner
and we confirm that the ammonia thermometer is robust.

\end{abstract}


\section{Introduction}

Ammonia (NH$_3$) was the first polyatomic molecule to be detected in
the interstellar medium (ISM) by \cite{Cheung:68}. NH$_3$ is an
ubiquitous molecule in the ISM and an invaluable probe of the physical
conditions thanks to its many observable transitions that are
sensitive to excitation conditions. NH$_3$ has been detected in a
large variety of interstellar environments, ranging from pre-stellar
cores to external galaxies \citep{ho:83}.

With the launch of the far-infrared Herschel Space Observatory in
2009, new opportunities have been created to perform observations of
NH$_3$, together with other related nitrogen hydrides such as NH and
NH$_2$. These observations have allowed a new insight into the
interstellar nitrogen chemistry. For exemple, \cite{legal:14} have
derived NH:NH$_2$:NH$_3$ abundance ratios of 3:1:19 towards the
external envelope of the protostar IRAS16293. Towards diffuse
interstellar clouds, \cite{persson:10} have obtained a similar
NH:NH$_2$ ratio of 2:1 but a much larger NH$_2$:NH$_3$ ratio of $\sim$
1:1. NH$_3$ also exhibits nuclear spin symmetry states whose relative
abundances have been measured in the cold diffuse gas with Herschel
\citep{persson:12}. The derived anomalous (non-thermal) NH$_3$
ortho-to-para ratio (OPR) of $\sim 0.6$ was shown to reflect the small
H$_2$ OPR ($\lesssim 10^{-3}$) in the gas phase \citep{faure13}.

Ammonia has always been recognized as a good thermometer
\citep{walmsley83} because the relative populations of the metastable
levels $j=k$ (where $j$ is the total angular momentum quantum number
of NH$_3$, and $k$ is the projection of $j$ on the $C_3$ rotational
axis) involved in the inversion transitions are mainly determined by
collisional processes. In addition, as these transitions arise from
quite different energy levels but take place in a narrow wavelength
window close to 1.3~cm, they can be probed by a single telescope and
receiver, i.e. at the same spectral and spatial resolution and with
small relative calibration errors. The calibration of the ammonia
thermometer however depends critically on the accuracy of the
collisional rate coefficients.

The collisional excitation of NH$_3$ has been the object of numerous
studies since the late 1960s. Following the pioneering double
resonance experiments on NH$_3$--He performed by \cite{oka:67}, the
first theoretical study of ammonia inelastic collisions with He was
investigated by \cite{Green:76top}. An electron gas approximation of
the potential energy surface (PES) was employed and the quantal
dynamical formalism was introduced for the first time for the
excitation of a symmetric top molecule by a spherical atom, with the
inclusion of the inversion motion. The first scattering calculations
for NH$_3$ in collision with para-H$_2(j_2=0)$ (hereafter the H$_2$
rotational states are denoted by $j_2$) were performed in the 1980s by
\cite{danby88} based on the PES of \cite{danby86}. The derived rate
coefficients were employed to recalibrate the ammonia thermometer
which was previously based on NH$_3$--He rate coefficient
calculations. The accuracy of the PES of \cite{danby86} was later
checked against pressure broadening measurements at low temperature
(12.5--40~K) and a good agreement was found between theory and
experiment \citep{willey:02}. It was also shown that broadening cross
sections involving para-H$_2(j_2=0)$ were up to four times larger than
those obtained with He.

Revised collisional rate coefficients for NH$_3$ colliding with He and
para-H$_2(j_2=0)$ were computed recently using improved PESs. The
NH$_3$--He PES of \cite{Hodges:01}, generated using scaled
perturbation theory calculations, was employed by \cite{machin:05} to
perform quantal close-coupling and approximate coupled-states
scattering calculations. Their cross sections were compared to the
crossed beam measurements of \cite{schleipen:91} at a relative kinetic
energy of 436~cm$^{-1}$. The agreement between the theoretical parity
averaged cross sections and the experimental data was found to be
satisfactory but significant discrepancies were observed for some
parity-resolved state-to-state cross sections. Rate coefficients were
obtained for kinetic temperatures ranging from 5 to 300~K and for
transitions up to $j=4$ for para-NH$_3$, and up to $j=7$ for
ortho-NH$_3$. An improved NH$_3$--He PES was also obtained more
recently by \cite{gubbels:12} using a coupled-cluster method, from
which cross sections (but no rate coefficients) were computed. A very
satisfactory agreement was obtained with this PES between theoretical
and experimental state-to-state differential cross sections (DCSs)
\citep{tkac:14}.

The revised NH$_3$--para-H$_2(j_2=0)$ rate coefficients were computed
by \cite{maret09} at the full close-coupling level using a new PES
computed at the coupled cluster level.  The NH$_3$--para-H$_2(j_2=0)$
rate coefficients were found in good agreement with the previous
calculations by \cite{danby88} (within a factor of 2) and significant
differences with the NH$_3$--He system were found. Rate coefficients
were deduced for transitions among ammonia levels with $j \le 3$ and
kinetic temperatures in the range 5-100~K. It should be noted that the
accuracy of the NH$_3$--H$_2$ PES of \cite{maret09} was recently
checked against various experiments.  In particular, state-to-state
resolved DCSs for the inelastic scattering of ND$_3$ with H$_2$ at a
collision energy of 580~cm$^{-1}$ were measured by \cite{tkac:15} and
satisfactorily reproduced by quantal calculations. The isotropic part
of the NH$_3$--H$_2$ PES was also probed by the molecular beam
experiment of \cite{pirani:13}, which has provided a further accuracy
assessment by measuring the glory quantum interference
structure. Finally, the rotational spectrum of the NH$_3$--H$_2$ van
der Waals complex was measured very recently by \cite{surin:17} and
the deduced rotational constants were found to agree within 1--2\%
with those predicted from the PES of \cite{maret09}. The overall
agreement between theory and experiment for the NH$_3$--He and
NH$_3$--H$_2$ systems is an important testimony of the good accuracy
of the most recent available PESs.

The story of ammonia excitation, however, is not over yet. Indeed,
none of the above NH$_3$--H$_2$ studies have considered so far the
rotational structure of H$_2$. In the calculations of both
\cite{danby88} and \cite{maret09}, H$_2$ was restricted to its
(spherically symmetrical) para $j_2=0$ state. As a result, there is no
rate coefficient available for collisions with ortho-H$_2$, even if
some ortho-H$_2$ cross sections were computed at a few selected
energies \citep{offer:89,Rist:93} with the PES of \cite{danby86}. In
addition, the calculations of \cite{maret09} for para-H$_2$ need to be
improved by including $j_2=2$ in the scattering basis set. Finally, to
the best of our knowledge, there is no collisional data available for
collisions of NH$_3$ with H, which are of great importance in
molecular gas with a low H$_2$ fraction.

In this paper, we provide new NH$_3$--H$_2$ calculations taking into
account the rotational (non-spherical) structure of H$_2$, using the
NH$_3$--H$_2$ PES of \cite{maret09}. Similar calculations were
recently performed for the NH$_3$ deuterated isotopologues
(i.e. NH$_2$D, NHD$_2$ and ND$_3$) by \cite{Daniel:14,Daniel:16}. The
data of \cite{maret09} are also extended to higher temperatures. In
addition, we provide the first set of rate coefficients for the
NH$_3$--H collisional system. For these calculations, we use the
accurate full-dimensional NH$_4$ PES of \cite{Guo:14} which was
recently employed in a NH$_2$--H$_2$ scattering study
\citep{Bouhafs:17}.
 
The paper is organized as follow: The PESs and the scattering
calculations are presented in Section~2. In section~3, we report
state-to-state cross sections and rate coefficients for the rotational
excitation of NH$_3$ by H and H$_2$. The consequences for the ammonia
thermometer are discussed in section~4. Concluding remarks are drawn
in Section~5.

\section{Methods}

\subsection{Potential energy surface}

The NH$_3$--H$_2$ PES was computed at the coupled cluster with single,
double, and perturbative triple excitations [CCSD(T)] level of theory
with a basis set extrapolation procedure, as described in
\cite{maret09} where full details can be found. The ammonia and
hydrogen molecules were both assumed to be rigid, which is justified
at kinetic temperatures lower than $\sim$1000~K since the lowest
vibrational mode of NH$_3$ opens at 950~cm$^{-1}$. The ammonia
``umbrella'' (inversion) motion was thus not described in the PES
which depends on five coordinates only: the intermolecular distance
$R$ between the centers of mass of NH$_3$ and H$_2$ and four
orientation angles.
The NH$_3$ and H$_2$ geometries were taken at their ground-state
averaged values: $r_{\rm NH}$=1.9512~$a_0$, $\widehat{\rm
  HNH}$=107.38$^{\circ}$ and $r_{\rm HH}$=1.4488~$a_0$. 
A total of 118,000 {\it ab initio} points were computed by
\cite{maret09}. These grid points were chosen for 29 intermolecular
distances (in the range 3--15~$a_0$) via random sampling for the
angular coordinates. At each intermolecular distance, the interaction
energy was least-squares fitted, using a 167-term expansion including
anisotropies up to $l_1$=10 and $l_2$=4, where the integer indices
$l_1$ and $l_2$ refer to the NH$_3$ and H$_2$ angular dependence,
respectively. These 167 expansion terms were selected using a Monte
Carlo error estimator defined in \cite{rist:11}. The root mean squared
error (RMSE) is lower than 1~cm$^{-1}$ for $R>5$~$a_0$. A cubic spline
interpolation was finally employed over the intermolecular distance
grid and it was smoothly connected with standard extrapolations to
provide continuous radial expansion coefficients. We note that this
fit is different from that performed by \cite{maret09} because the
potential expansion is expressed here in the body-fixed coordinates
adapted to the \texttt{HIBRIDON} scattering code, as in \cite{ma:15},
whereas the \texttt{MOLSCAT} scattering code \citep{molscat:94} was
employed by \cite{maret09}. The global minimum deduced from our fit,
lies at -267~cm$^{-1}$ for $R=6.1$~$a_0$, with H$_2$ colinear with the
$C_{3}$ axis of ammonia at the nitrogen end. This result can be
compared with other recent calculations: \cite{mladenovic:08} and
\cite{sheppleman:12} found the global minimum at a similar location
but with smaller binding energies of -253~cm$^{-1}$ and
-245~cm$^{-1}$, respectively.
 
The NH$_3$--H PES was constructed from the recently computed
nine-dimensional global PES of the ground electronic state of NH$_4$,
as described by \cite{Guo:14} where full details can be found. This
global PES was determined at the explicitely correlated unrestricted
coupled-cluster level of theory using an augmented
correlation-consistent triple zeta basis set
(UCCSD(T)-F12a/aug-cc-pVTZ). The abstraction and exchange channels,
respectively NH$_3$+H $\leftrightarrow$ NH$_2$ + H$_2$ and NH$_3$+H
$\leftrightarrow$ NH$_4 \leftrightarrow$ NH$_3$ + H, are thus included
in the PES. They can be safely neglected here, because they involve
energy barriers higher than 3200~cm$^{-1}$, which is much higher than
the investigated collision energies\footnote{The abstraction reaction
  is slow, with a rate coefficient lower than
  10$^{-15}$~cm$^3$s$^{-1}$ at temperatures lower than 500~K
  \citep{Guo:14}}. About 100,000 {\it ab initio} points were computed
by \cite{Guo:14} and they were fitted using the permutation-invariant
polynomial neutral network (PIP-NN) method with a RMSE of
27~cm$^{-1}$. The RMSE of 27~cm$^{-1}$ is for the full
nine-dimensional PES (including the reactive path).  The fitting RMSE
for the non-reactive NH$_3$--H region relevant for this work is much
smaller, of the order of a few cm$^{-1}$.  In this work, NH$_3$ is
considered as rigid with the ground-state averaged geometry of ammonia
given above. The NH$_3$--H PES is thus described as a function of
three coordinates: the distance $R$ between the center of mass of
NH$_3$ and the H atom, and two spherical angles.
As the original routine of \cite{Guo:14} employs internuclear
coordinates, the spherical to cartesian transformation was employed to
determine the cartesian positions of the H atom in the ammonia
body-fixed frame (see \cite{Bouhafs:17} for the general transformation
in NH$_2$-H$_2$). The original fit of \cite{Guo:14} was employed to
generate interaction energies on a dense grid of 90,000 geometries,
chosen for 30 distances $R$ (in the range 3--20~$a_0$) via random
sampling for the angular coordinates. An asymptotic potential of
228.297~cm$^{-1}$ (corresponding to the above monomer averaged
geometries) was subtracted from these interaction energies. At each
intermolecular distance, the interaction energy was least-squares
fitted, using a 26-term expansion including all anisotropies up to
$l_1$=10. The RMSE is lower than 1~cm$^{-1}$ for $R>4$~$a_0$. The
global minimum, as deduced from our fit, lies at -78.31~cm$^{-1}$ for
$R=6.2$~$a_0$, with the H atom close to the equatorial location
(16$^{\circ}$ below) and equidistant from the two closest H atoms of
ammonia. Similar equilibrium geometries were found for the minima in the NH$_3$--He
and NH$_3$--para-H$_2(j_2=0)$ PES, with corresponding energies of
-35.08~cm$^{-1}$ \citep{gubbels:12} and -85.7~cm$^{-1}$
\citep{maret09}, respectively.

\subsection{Scattering calculations}

The rotation-inversion levels of the NH$_3$ symmetric top molecules
are labeled as $j_k^{\epsilon}$, where $j$ is the total angular
momentum quantum number of the molecule, $k$ is the projection of $j$
on the $C_3$ axis and $\epsilon =\pm $ is the symmetry index. For each
rotational state $j_k^{\epsilon}$, the umbrella inversion symmetry is
equal to $-(-1)^j\epsilon$ and the parity index of the
rotation-inversion wave function is $-\epsilon (-1)^{j+k}$
\citep{Rist:93}. Because of the symmetry under permutation of the
three identical protons, the rotational states are split into two $k$
stacks~: ortho-NH$_3$ which correspond to $k$=$3n$ (with $n$ an
integer) and para-NH$_3$ with $k\neq 3n$ (hereafter denoted as
o-NH$_3$ and p-NH$_3$ respectively). As
for NH$_3$ the rotational states of H$_2$ molecule split into para and ortho
nuclear-spin permutation symmetry modifications. The para states of
H$_2$ have even rotational states $j_2 = 0, 2, ... $ and the ortho
states have odd rotational states, $j_2 = 1, 3, ... $ (hereafter
p-H$_2$ and o-H$_2$, respectively).

Since the umbrella inversion motion of the NH$_3$ molecule is not
described in the PES (see above), the inversion-tunnelling
wavefunctions were approximated as even and odd combination of the
two rigid structures, as first suggested by
\cite{Green:76top}. Previous studies of the inelastic scattering of
NH$_3$ by He and Ar atoms have demonstrated that this approximation
is in very good agreement (i.e. better than 10\%) with an elaborate
treatment of the umbrella motion \citep[e.g.][]{gubbels:12}.

We used the rotational constants $A$=$B$=9.9402 cm$^{-1}$ and
$C$=6.3044 cm$^{-1}$ for NH$_3$ and $B_0$=59.3801 cm$^{-1}$ for H$_2$.
The experimental ground state inversion splitting of 0.7903~cm$^{-1}$
was also included. The reduced masses of the NH$_3$--H$_2$ and
NH$_3$--H systems are 1.802 and 0.952 amu, respectively.

As the ortho and para levels of NH$_3$ and H$_2$ do not interconvert
in inelastic collisions, these were treated separately. All
calculations were performed using the \texttt{HIBRIDON} scattering
code.  The scattering equations were solved using the almost exact
close-coupling approach.  In the case of NH$_3$--H$_2$, the 167-term
PES expansion (including anisotropies up to $l_1$=10 and $l_2$=4) was
reduced in order to save computational time. In the close-coupling
calculations, we thus adopted a $55$-terms expansion including
anisotropies up to $l_1$=6 and $l_2$=2. In the case of NH$_3$--H, all
26 terms of the PES expansion were used.

Scattering calculations were performed for total energies up to
1600~cm$^{-1}$ for NH$_3$-H$_2$ and up to 2000~cm$^{-1}$ for
NH$_3$-H. We computed inelastic cross sections between levels with an
internal energy lower than 419~\cm, that is up to $j_k$=$6_0$ for
o-NH$_3$ and $j_k$=$6_1$ for p-NH$_3$. For collisions with p-H$_2$,
the $j_2$=0 and $j_2$=2 levels of H$_2$ are included in the basis set,
while for o-H$_2$ collisions, only the $j_2$=1 level was retained.
For both collisional partners, the NH$_3$ rotational basis set also
included several closed (or higher than $419$ cm$^{-1}$) energy
levels. Thus, at the highest investigated total energies and for
collisions with H$_2$, the basis set included all levels with $j \leq$
10 and 12 for o- and p-NH$_3$, respectively. For collisions with H,
the NH$_3$ basis set included all levels with $j \leq$ 9 for o- and
p-NH$_3$.  These NH$_3$ and H$_2$ rotational basis sets allow the
determination of cross sections converged within $10\%$ accuracy.  The
integration parameters were also chosen to ensure convergence of the
cross sections within a few percent. An energy step of 0.1 cm$^{-1}$
was used at low energies to properly describe resonances. This energy
step has been progressively increased with increasing collision
energy.

Finally, rate coefficients were obtained by integrating the cross
sections over a Maxwell-Boltzmann distribution of relative velocities:  
\begin{eqnarray}
\label{thermal_average}
k_{\alpha \rightarrow \beta}(T) & = & \left(\frac{8}{\pi\mu k^3_{B} T^3}\right)^{\frac{1}{2}} \nonumber\\
&  & \times  \int_{0}^{\infty} \sigma_{\alpha \rightarrow \beta}\, E_{c}\, e^{-\frac{E_c}{k_{B}T}}\, dE_{c}
\end{eqnarray}
\noindent where $\sigma_{\alpha \rightarrow \beta}$ is the cross
section from the initial state $\alpha$ to the final state $\beta$,
$\mu$ is the reduced mass of the system and $k_{B}$ is the Boltzmann's
constant. The range of total energies mentioned above allowed us to
determine rate coefficients up to 200~K.

\section{Results}

\subsection{Cross sections}

\begin{figure}
\centering{\includegraphics[width=8.5cm]{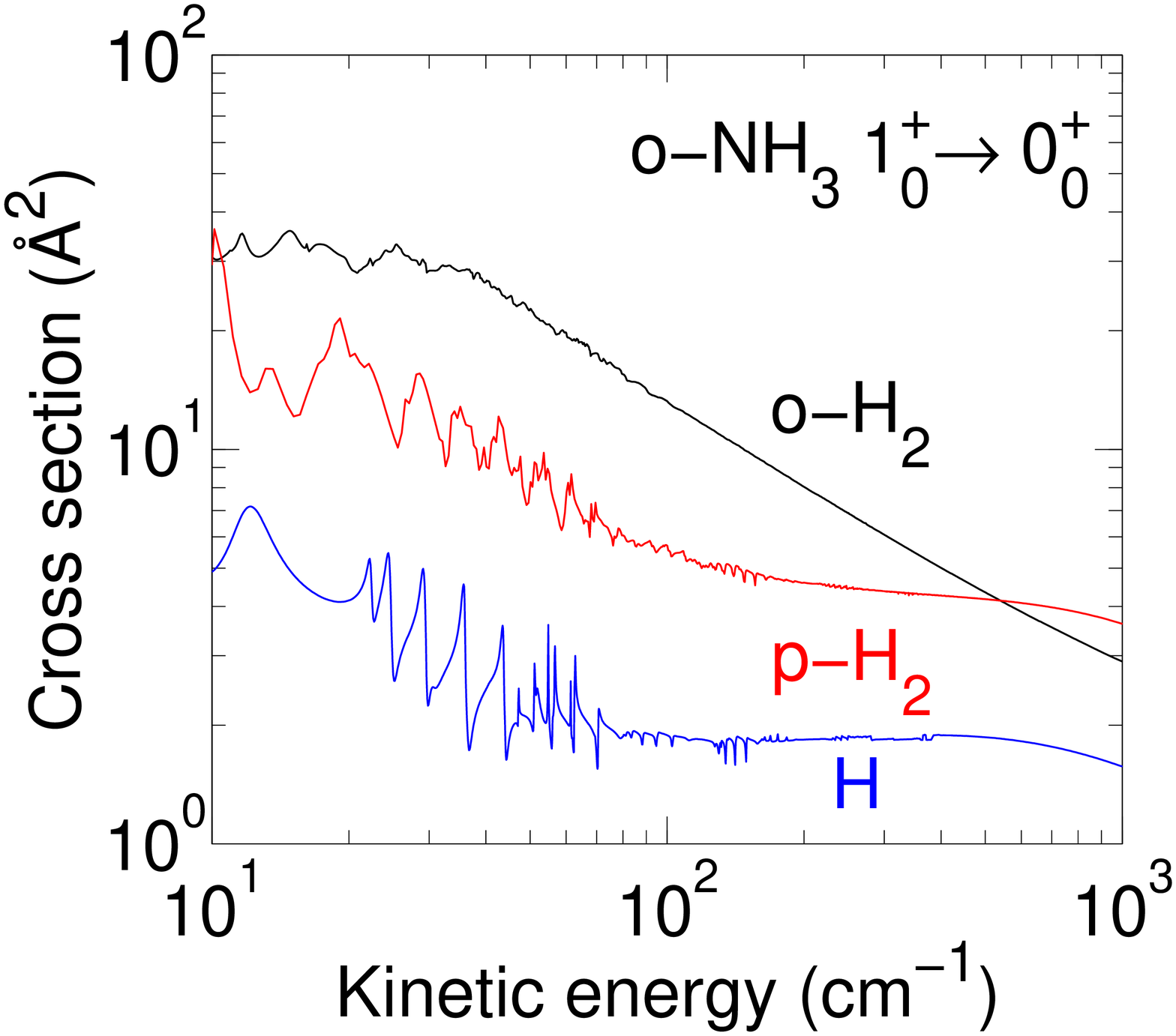}}
\centering{\includegraphics[width=8.5cm]{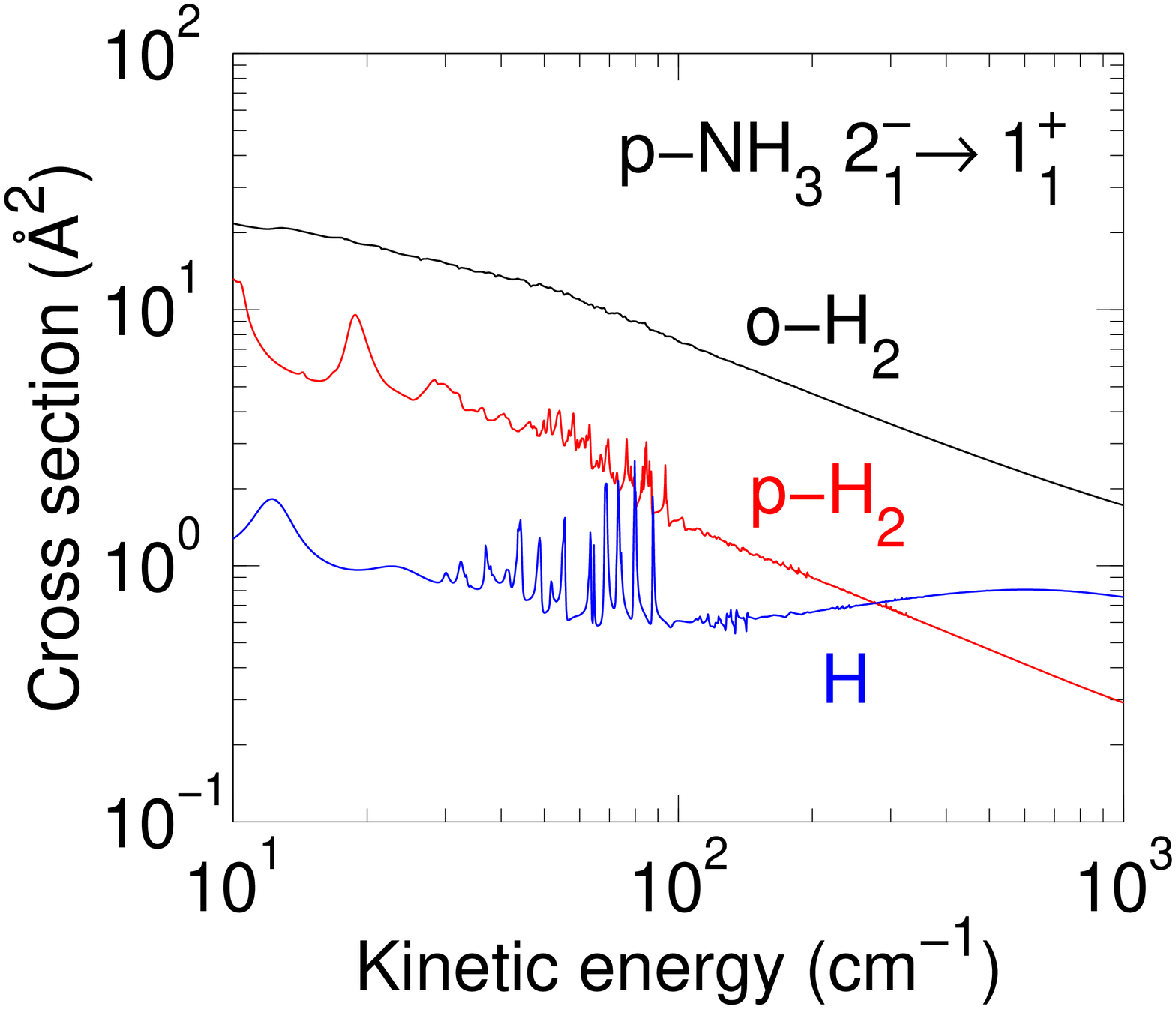}}
\caption{Rotational de-excitation cross sections of o- and p-NH$_3$ by
  p-H$_2( j_2=0$) and o-H$_2({j_2}=1$) and H for the transitions
  $1_0^+\rightarrow 0_0^+$ and $2_1^-\rightarrow 1_1^+$.}\label{fig:1}
\end{figure}

Figure~\ref{fig:1} shows the collisional energy dependence of the
$1_0^{+} \rightarrow 0_0^{+}$ and $2_1^{-} \rightarrow 1_1^{+}$
de-excitation cross sections of o- and p-NH$_3$, respectively, in
collision with p-H$_2(j_2 = 0$), o-H$_2(j_2 = 1$) and H.

Important resonance peaks appear with both H and H$_2$ colliders. This
is related to the presence of an attractive potential well, which
allows for bound and quasi-bound states to be formed before the
complex dissociates. Both shape and Feshbach resonances are then
expected \citep{ma:15}.

The cross sections for collisions with o-H$_2$($j_2 =1$) display
smoother variations because of resonance overlaps. Indeed, for o-H$_2$
we expect a large number of quasi-bound states due 1to the contribution
of an additional coupling terms $j$+$j_2$=$j_{12}$ absent in
collisions with p-H$_2(j_2 =0$).

At low to intermediate energies, collisional cross sections with
H$_2(j_2 =0,1$) are larger than that those with H. This reflects the
deeper potential well of the NH$_3$--H$_2$ PES compared to that of the
NH$_3$--H PES. Similarly, the NH$_3$--o-H$_2(j_2=1)$ cross sections
are larger than the NH$_3$--p-H$_2(j_2=0)$ because of the permanent
quadrupole moment of H$_2$, which vanishes for $j_2 = 0$ but not for
$j_2 = 1$, leading to a stronger interaction of NH$_3$ with
o-H$_2(j_2=1)$ than with p-H$_2(j_2=0)$.

As mentioned above, \cite{ma:15} have also employed the PES of
\cite{maret09} to compute inelastic cross sections for
NH$_3$--H$_2$. We have compared our new results with those obtained by
\cite{ma:15} and both sets of cross sections were found to agree
within a few percent, as expected.
 
\subsection{Rate coefficients}

\begin{figure}
\centering{\includegraphics[width=8.5cm]{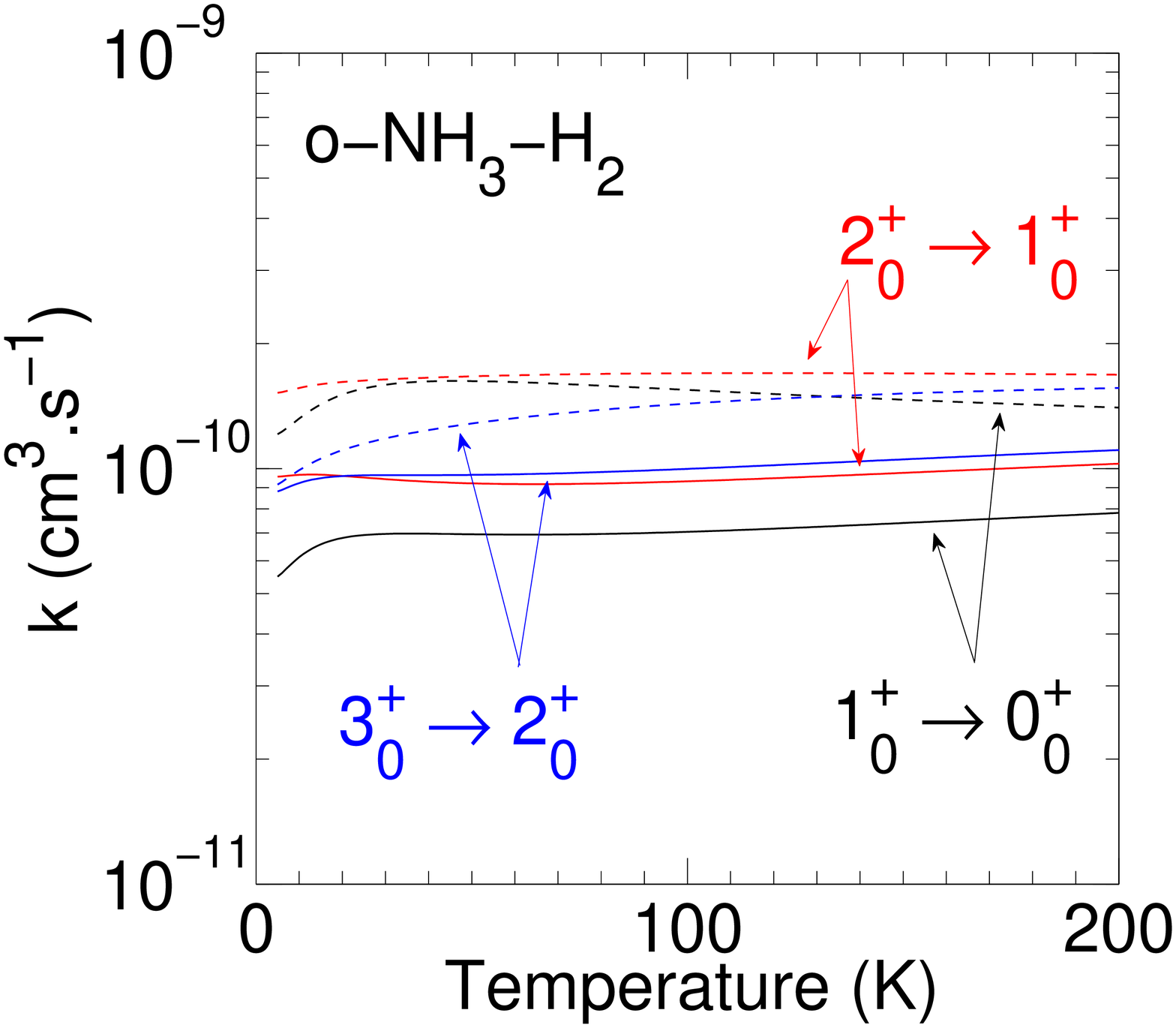}}
\centering{\includegraphics[width=8.5cm]{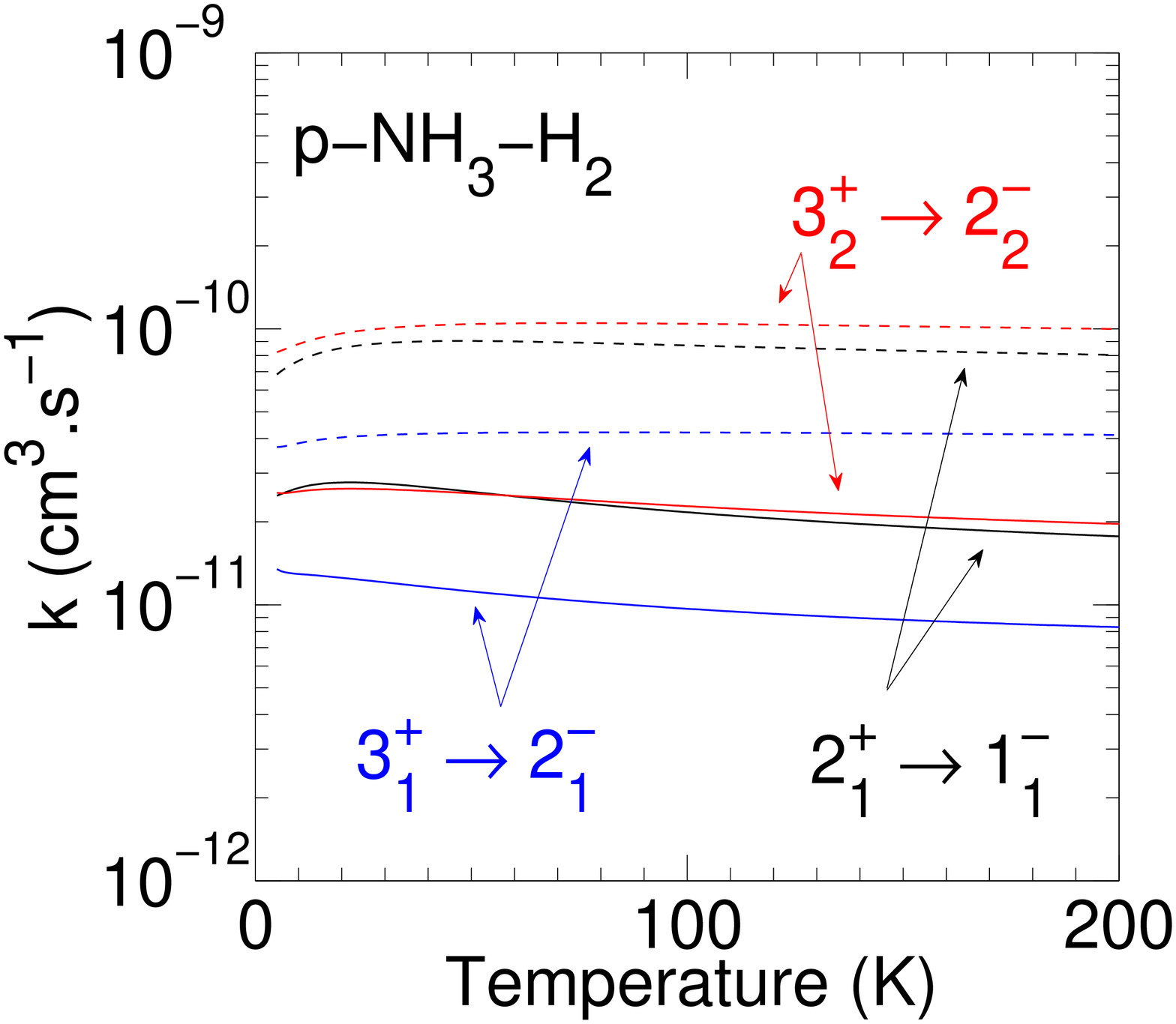}}
\caption{Thermal dependence of the rate coefficients of o- and
  p-NH$_3$ by p-H$_2$( j$_2$=0) (solid line) and o-H$_2({j_2}=1)$
  (dashed line) for a set of transitions with $\Delta
  j=1$.}\label{fig:2}
\end{figure}
\begin{figure}
\centering{\includegraphics[width=8.5cm]{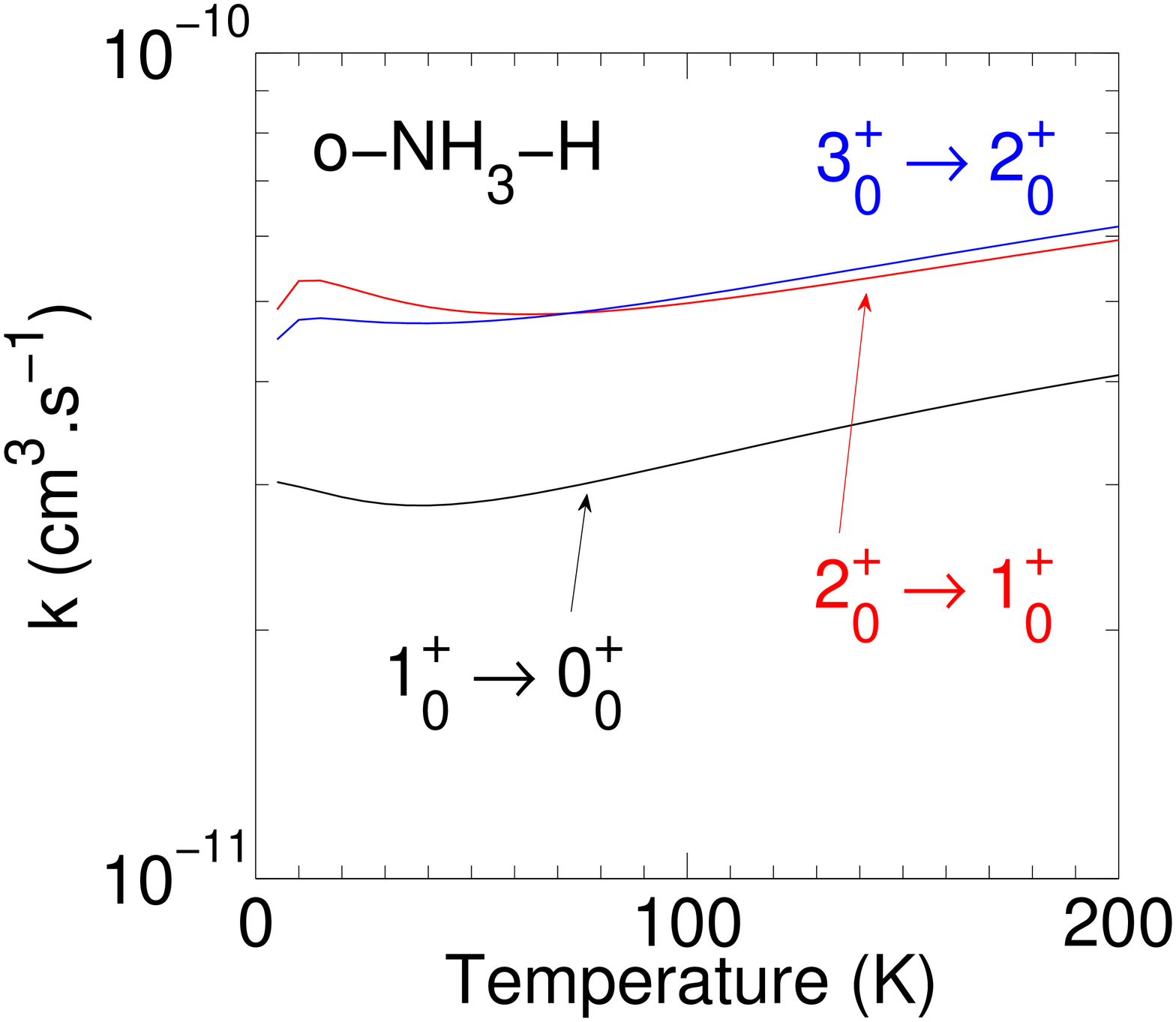}}
\centering{\includegraphics[width=8.5cm]{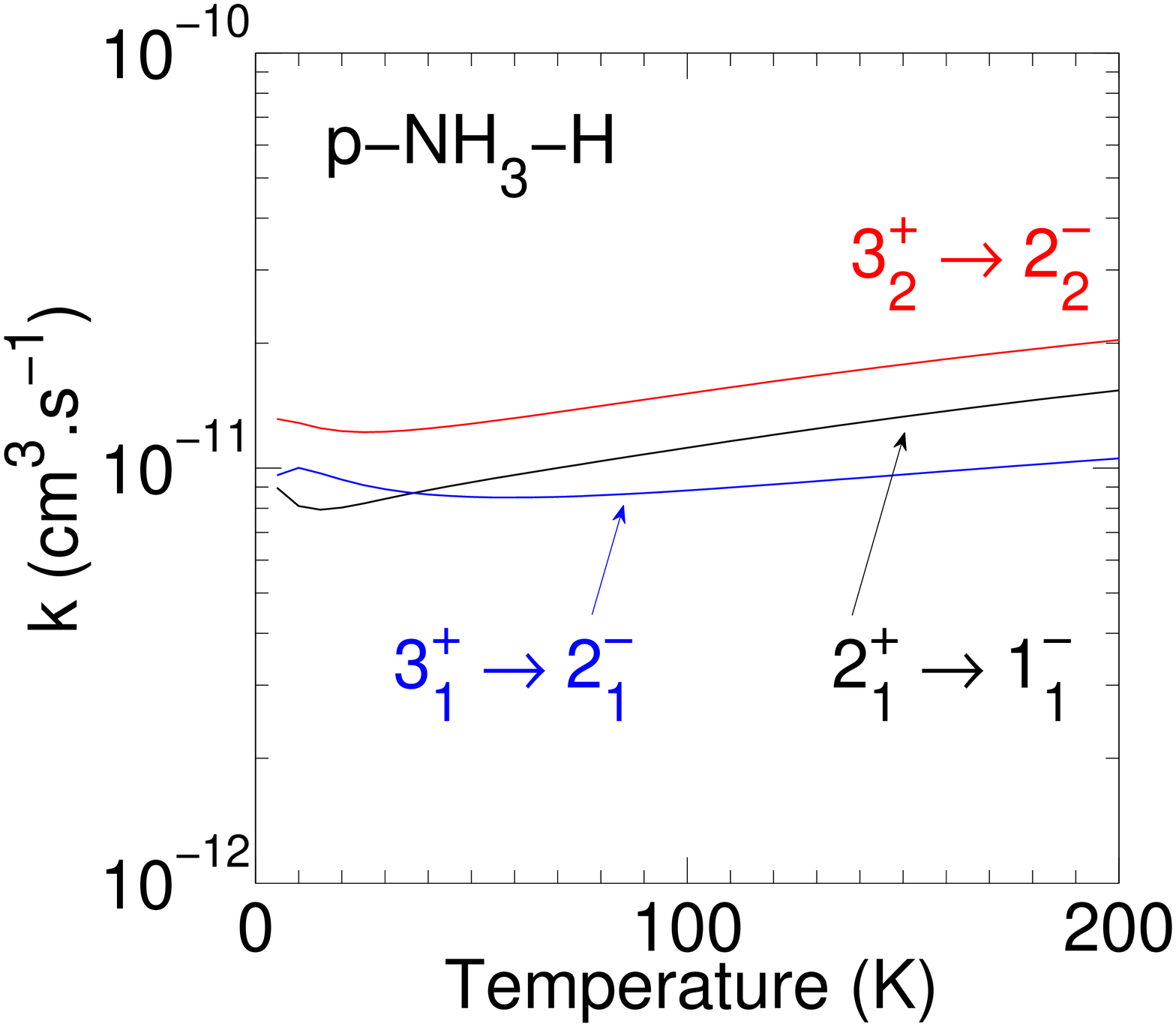}}
\caption{Thermal dependence of the rate coefficients of o- and
  p-NH$_3$ by H for a set of transitions with $\Delta
  j=1$.}\label{fig:3}
\end{figure}

In Figs. \ref{fig:2} and \ref{fig:3}, de-excitation rate coefficients
are presented for collisions of o- and p-NH$_3$ with H$_2$ and H.

We observe a smooth temperature variation of the de-excitation rate
coefficients that slightly increase with increasing temperature.  As
expected from the above discussion, the rate coefficients for
collisions with o-H$_2(j_2 = 1)$ are larger than those for collisions
with p-H$_2(j_2 = 0)$.

\begin{figure}
\centering{\includegraphics[width=8.5cm]{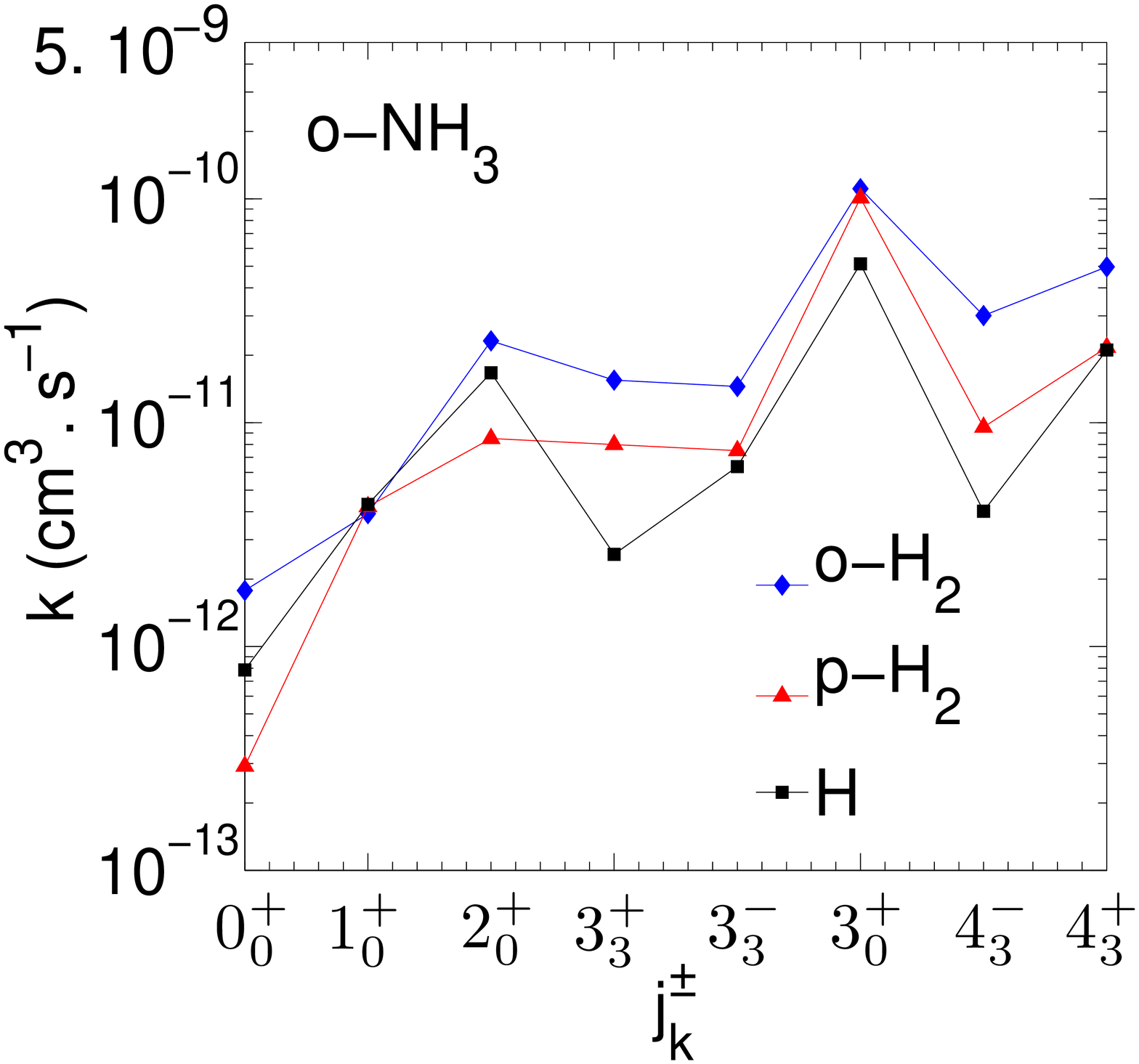}}
\centering{\includegraphics[width=8.5cm]{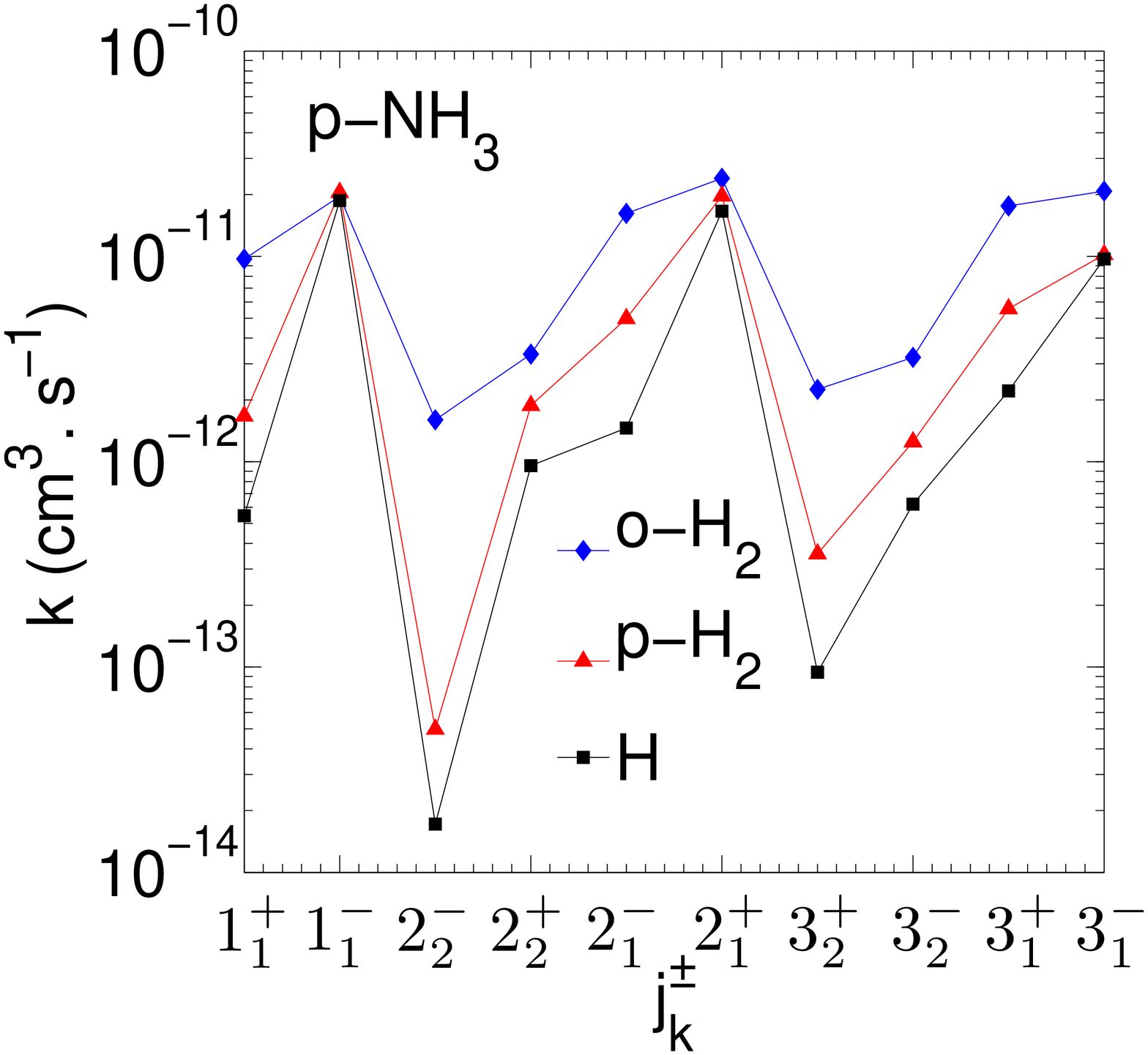}}
\caption{Rate coefficients for transitions out of the initial NH$_3$
  states $j_k^{\epsilon}$ =$4_0^+$ (upper panel) and $j_k^{\epsilon}$
  =$4_4^+$ (lower panel) due to collisions with p-H$_2( j_2$=0),
  o-H$_2({j_2}=1)$ and H at $T$=100~K.}\label{fig:4}
\end{figure}

We now examine the collisional propensity rules.
Figure~\ref{fig:4} shows, at 100~K, the rotational de-excitation rate
coefficients for transitions out of o-NH$_3$($j_k^{\epsilon}$
=$4_0^+$) and p-NH$_3$($j_k^{\epsilon}$ =$4_4^+$) due to collisions
with H$_2(j_2 = 0,1)$ and H. For o-NH$_3$ in its $4_0^+$state (upper
panel), we notice a marked de-excitation propensity rules in favor of
the $3_0^+$ state ($\Delta j = 1$, $\Delta k=0$ and symmetry
conserving) and in favor of the $4_3^+$ state ($\Delta j=0$, $\Delta
k=3$, and symmmetry conserving). For p-NH$_3$ in its $4_4^+$state
(lower panel), the preferred transitions correspond to $\Delta k=3$,
since there is no de-excitation with $\Delta k=0$ for this initial
level. No propensity on the symmetry (or parity) index is observed. As
can be seen from Fig.~\ref{fig:4}, the same general trends are
observed for the three different colliders p-H$_2(j_2 = 0)$,
o-H$_2(j_2 = 1$) and H, although propensities tend to weaken in the
case of o-H$_2$($j_2 = 1$) due to the additional angular couplings. To
summarize, collisional transitions are found to favor transitions with
$\Delta j=0, \pm 1, \pm 2$ and $\Delta k=0, \pm 3$ with no clear
propensity on the symmetry (or parity) index. This can be compared to
the radiative (dipolar) selection rules which obey $\Delta j=0, \pm
1$, $\Delta k=0$ and parity changing.

\subsection{Comparison with previous data}

In astrophysical applications, it is usual to infer the rate
coefficients of a colliding system from the values calculated for a
closely related system (i.e. same target but with different
collisional partners), using a scaling factor coefficient such as the
square root of the reduced mass ratio (i.e. assuming identical cross
sections for the two colliders).

\begin{figure}
\centering{\includegraphics[width=8.5cm]{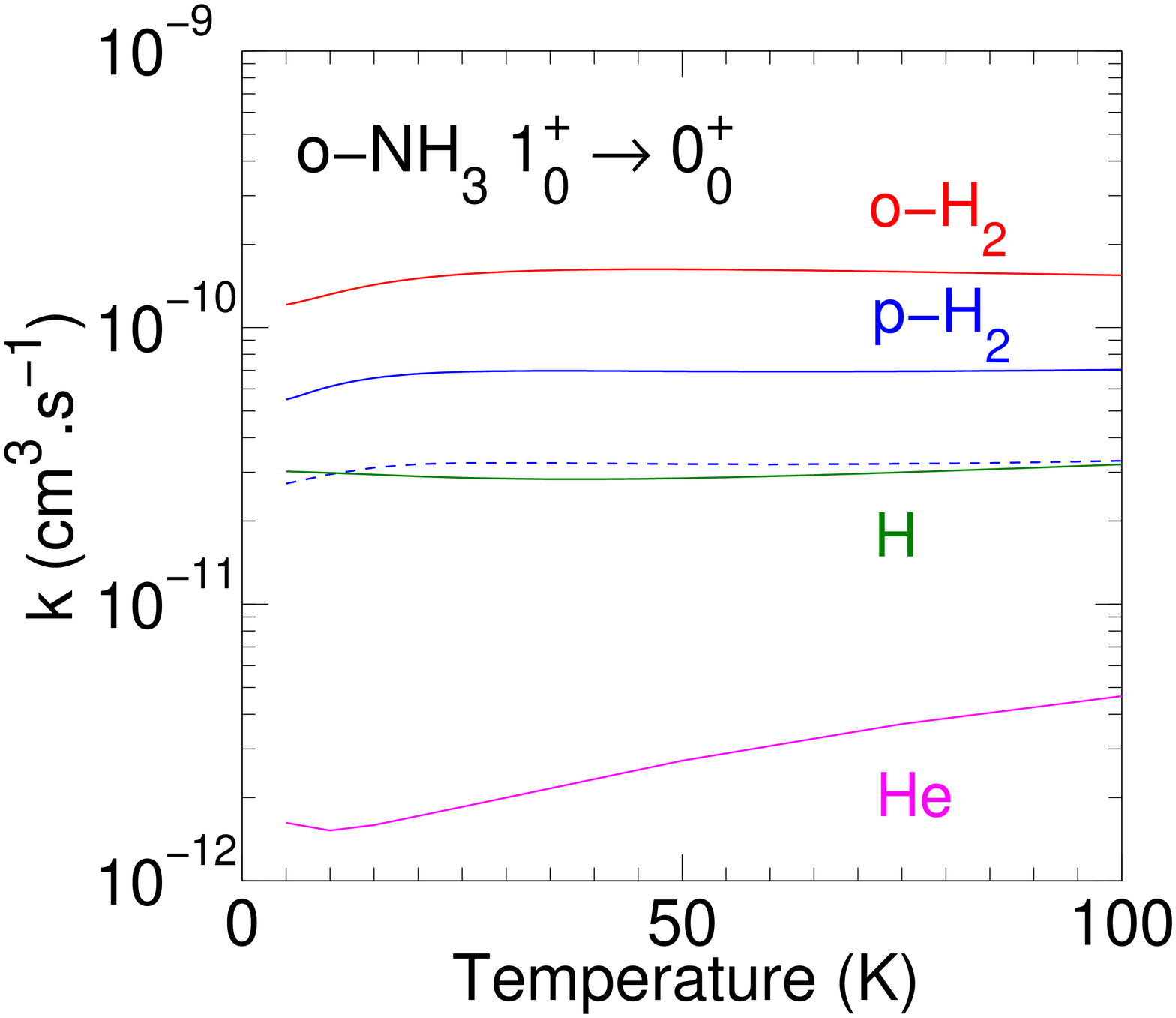}}
\centering{\includegraphics[width=8.5cm]{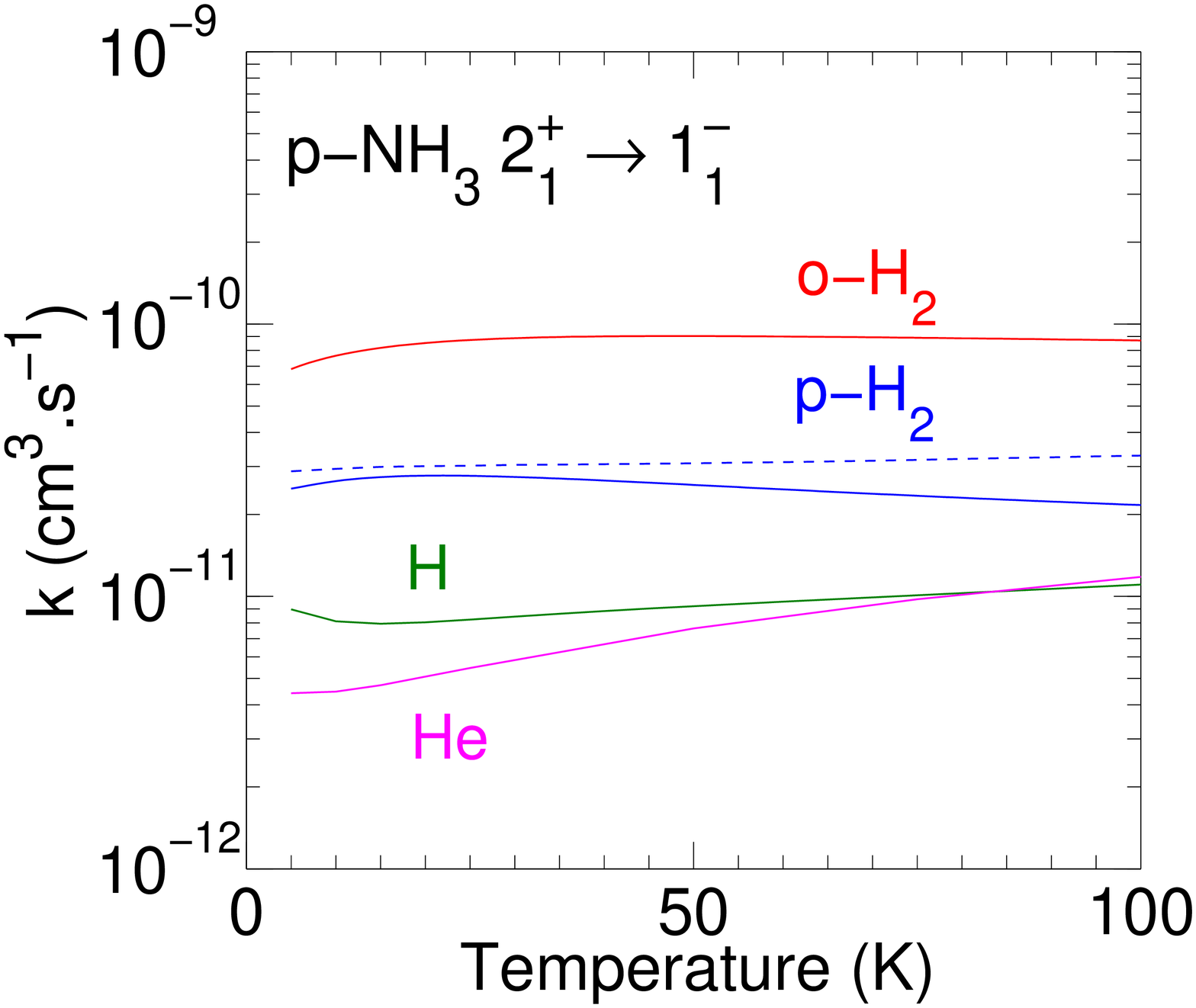}}
\caption{Comparison between the rate coefficients for collisional
  de-excitation of o- and p-NH$_3$ by H, o-H$_2$ and p-H$_2$ (dashed
  line for the rate coefficients of Maret et al. (2009)) and He (from
  Machin et al. (2005)). }\label{fig:6}
\end{figure}

In order to assess the validity of such a simple scaling law, we
compare in Fig.~\ref{fig:6} the current rate coefficients of both o-
and p-NH$_3$ in collision with H$_2$, H and those obtained by
\cite{maret09} and \cite{machin:05} for the spherical p-H$_2(j_2=0)$
and He colliders, respectively. Assuming the (reduced mass) scaling
law is valid, the rate coefficients for p-H$_2(j_2=0)$ and H should be
about 50\% and 80\% larger than the He rate coefficients,
respectively, owing to the smaller collisional reduced
masses. However, as expected and found for many others hydrides like
HCl \citep{lanza:14}, H$_2$O \citep{daniel:06}, etc. it can be seen in
Fig.~\ref{fig:6} that the rate coefficient ratios vary significantly
with both the temperature and the transition considered. The rate
coefficients for H$_2$ (both o- and p-H$_2$) are up to two orders of
magnitude larger than those for He \citep{machin:05}. The rate
coefficients for collisions with H are not larger than those with
p-H$_2(j_2=0)$, contrary to what could be anticipated from the
(reduced mass) scaling law. The differences between the three
colliders are most pronounced at low temperature and can be explained
by the fact that at low collisional energy, cross sections are very
sensitive to the well depth of the PES. Here, the well depth is 3 and
8 times larger for H$_2$ than for H and He (-267~cm$^{-1}$
vs. -78.31~cm$^{-1}$ and -32.85~cm$^{-1}$, respectively). As a result,
we confirm that, for hydrides, He atoms are very poor substitutes for
H or H$_2$.

Regarding the rate coefficients for p-H$_2$, we can notice significant
differences (up to a factor of 2) between the present calculations and
those of \cite{maret09} (see Fig.\ref{fig:6}, upper panel). We recall
that these differences reflect the inclusion of the rotational
(non-spherical) structure of p-H$_2$ in our treatment. However, on the
average, the differences are less than 30\% and the temperature
variation of the two sets of data is quite similar. Hence, we do not
expect a major impact of the new p-H$_2$($j_2$=0) rate coefficients on
the astrophysical modelling. Additionally, as pointed out in our
previous figures, we observe again in Fig.~\ref{fig:6} a large
difference between p- and o-H$_2$. It is thus crucial to consider the
two nuclear spin species of H$_2$ as two distinct colliders.

The whole set of rotation-inversion rate coefficients for
p-H$_2$($j_2=0$), o-H$_2$($j_2=1$) and H will be available through the
LAMDA \citep{schoier:05} and BASECOL \citep{dubernet:13} data
bases. We note that in some environments the fraction of
p-H$_2$($j_2=2$) can become significant, e.g.  at kinetic temperatures
above $\sim$ 100~K. We have found that cross sections for H$_2$
initially in $j_2=2$ differ by less than 30\% with those for H$_2$ in
$j_2=1$ (see also \cite{Daniel:14} for NH$_2$D). As a result, it can
be assumed that rate coefficients for p-H$_2$($j_2=2$) are identical
to those for o-H$_2$($j_2=1$).

\section{The ammonia thermometer}

Because radiative dipolar transitions with $\Delta k\neq 0$ are
forbidden in NH$_3$ (except for very slow $\Delta k=\pm 3$
transitions), the exchange of population between the different
$k$-ladders occurs essentially via collisional processes. For many
years, the inversion transitions for different $k$-ladders at $\sim$
20-30~GHz have been employed to derive kinetic temperatures in both
cold ($<40$~K) and warm ($>100$~K) galatic and extra-galactic star
formation environments. The $(1_1^-\rightarrow
1_1^+)/(2_2^+\rightarrow 2_2^-)$ line ratio, in particular, has been
widely used to monitor the lowest kinetic temperatures ($<
40$~K). Considering only the first three doublets $1_1^\pm$, $2_2^\pm$
and $2_1^\pm$ and assuming that the population of the $2_1^\pm$
doublet is negligible relative to that in $1_1^\pm$, \cite{walmsley83}
have shown that the excitation temperature between the two lowest
doublets is given by the analytical formula:
\begin{equation}
  T_{1,2}^A=T_k\left\{1+\frac{T_k}{T_0}\ln\left[1+\frac{k(2_2\rightarrow 2_1)}{k(2_2\rightarrow 1_1)}\right]\right\}^{-1}
\end{equation}
where $T_0$ is the energy difference between the first two metastable
doublets ($41.2$~K), $T_k$ is the kinetic temperature and
$k(2_2\rightarrow 2_1)$ and $k(2_2\rightarrow 1_1)$ are the rotational
rate coefficients (i.e. averaged and summed over the inversion
symmetry index). In practice, this formula is accurate at temperatures
below 40~K and at densities lower than $10^5$~cm$^{-3}$. We note that
analytic expressions can be derived also for higher metastable levels
\citep{walmsley83}.

Observationally, $T_{1, 2}$ is determined by observing the hyperfine
components of the $1_1^-\rightarrow 1_1^+$ and $2_2^+\rightarrow
2_2^-$ inversion transitions. The inversion doublets have indeed
hyperfine components due to (mainly) the quadrupole moment of the
$^{14}$N nucleus. Asssuming the excitation temperature of each
hyperfine transition (within a doublet) is the same, one can derive
the opacities $\tau(1_1)$ and $\tau(2_2$) from which the excitation
temperature $T_{1, 2}$ can be obtained \citep{ho79,juvela12}.
In order to derive the kinetic temperature from the measured $T_{1,
  2}$ excitation temperature, one needs to calibrate the ammonia
thermometer, Eq.~(2), and a very good knowledge of the collisional
rate coefficients is necessary.

\cite{maret09} have investigated the robustness of the ammonia
thermometer by comparing the excitation temperature $T_{1, 2}$
computed using their rate coefficients with that computed using the
older collisional data of \cite{danby88}. Although the
rotation-inversion rate coefficients differ by up to a factor of 2,
they found almost identical excitation temperatures $T_{1, 2}$ (see
their Fig.~5) and concluded that the rate coefficients of
\cite{danby88} were of sufficient accuracy to calibrate the
thermometer.

\begin{figure}
\centering{\includegraphics[width=7.5cm,angle=-90]{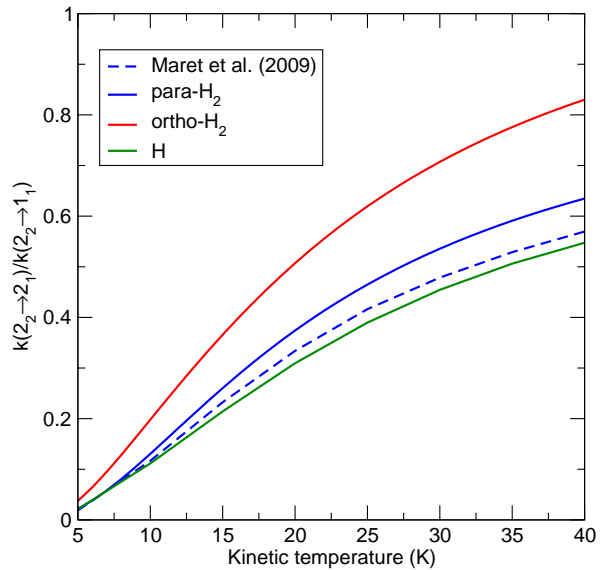}}
\caption{Ratio of the rate coefficients $k(2_2\rightarrow 2_1)$ and
  $k(2_2\rightarrow 1_1)$ as function of the kinetic
  temperature.}\label{fig:ratio}
\end{figure}

Here we test for the first time the impact of including the H$_2$
rotational structure on the ammonia thermometer. In
Fig.~\ref{fig:ratio}, the ratio of the rate coefficients
$k(2_2\rightarrow 2_1)$ and $k(2_2\rightarrow 1_1)$ is plotted as
function of temperature. We first observe that this ratio is lower
than unity whatever the collider. We notice that the inclusion of
$j_2=2$ in the H$_2$ basis set (neglected in \cite{maret09}) tends to
increase the ratio by up to 10\%. The difference between o- and
p-H$_2$ rate coefficients is even more significant, with up to a 30\%
increase of the ratio at 40~K. Finally, the collisions with hydrogen
atoms give the lowest ratio. These differences reflect small
variations in propensity rules, as illustrated in Fig.~\ref{fig:4}.

\begin{figure}
\centering{\includegraphics[width=7.5cm,angle=-90]{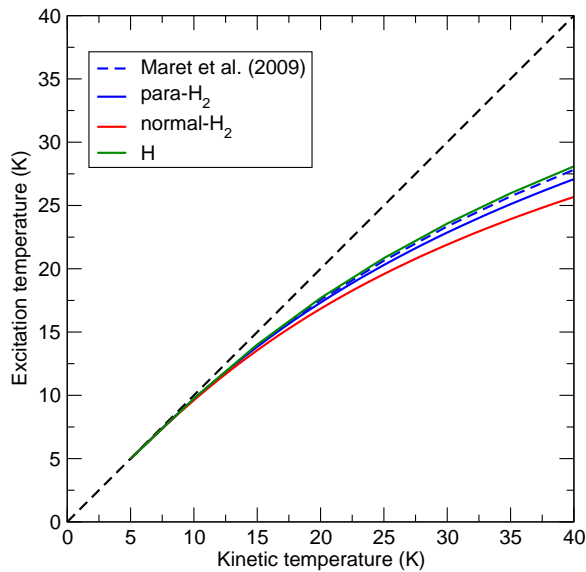}}
\caption{Excitation temperature between the $1_1$ and $2_2$ metastable
  levels of p-NH$_3$ as function of the kinetic temperature. The
  analytic expression, Eq.~(1), is employed with the rate coefficients
  computed in this study and from the rate coefficients of Maret et
  al. (2009) for p-H$_2$. The dashed straight line corresponds to $T_k = T_{1, 2}$.}\label{fig:thermo}
\end{figure}

The different ratios plotted in Fig.~\ref{fig:ratio} translate into
different excitation temperatures $T_{1, 2}$, as shown in
Fig.~\ref{fig:thermo}. In this plot, we can first notice that for
small kinetic temperature ($<$10~K), e.g. in cold prestellar cores,
$T_{1, 2}=T_k$ and $T_{1, 2}$ can be used as a direct proxy for the
kinetic temperature. Deviations between $T_{1, 2}$ and $T_k$ occur
above 10~K. As expected, the difference between the present results
for p-H$_2$ and those of \cite{maret09} is small and does not exceed
1~K. Indeed the logarithm in Eq.~(1) further reduces the differences
observed in Fig.~\ref{fig:ratio}. We also provide the excitation
temperatures $T_{1, 2}$ for a normal-H$_2$ gas, i.e. with an H$_2$ OPR
of 3, and for a purely atomic gas. The differences with p-H$_2$ are
found to be less than 2~K, suggesting that the ammonia thermometer
cannot be employed to provide indirect constraints on the H$_2$ OPR or
on the atomic fraction of the gas. Indeed, the three calibration
curves agree within 2~K and any estimate of the relative abundance of
o-H$_2$, p-H$_2$ and H from the measured $T_{1, 2}$ would require the
knowledge of $T_k$ with a sub-Kelvin accuracy, which is hardly
achievable with the current precision of observations.

Finally, in order to provide a useful calibration of the ammonia
thermometer for dark clouds and starless cores, we fitted the ratio of
the rate coefficients $k(2_2\rightarrow 2_1)$ and $k(2_2\rightarrow
1_1)$ in the case of p-H$_2(j_2=0)$. Indeed, although H$_2$ cannot
be directly observed at low temperature, indirect observations
indicate that the cold interstellar gas is mainly composed by
p-H$_2(j_2=0)$, with an OPR lower than 1\% below $\sim$30~K
\citep{troscompt09,pagani09,faure13}. From the results plotted in
Fig.~\ref{fig:ratio} we have derived:
\begin{equation}
  \frac{k(2_2\rightarrow 2_1)}{k(2_2\rightarrow 1_1)}=1.073\exp(-21.01/T_k),
\end{equation}
which has an accuracy better than 2\% above 10~K (remember $T_k=T_{1,
  2}$ below 10~K). This fit can be used to relate the kinetic
temperature $T_k$ to $T_{1, 2}$ {\it via} Eq.~(2), provided that
$T_k<40$~K. We also fitted the reverse relation and obtained:
\begin{equation}
T_k=T_{1, 2}\left\{1-\frac{T_{1, 2}}{T_0}\ln\left[1+1.608\exp(-25.25/T_{1, 2})\right] \right\}^{-1},
\end{equation}
which has an accuracy better than 0.3\%. We note that this analytical
formula was suggested by \cite{tafalla04}[see their Appendix B]. Their
numerical expression was based on the rate coefficients of
\cite{danby88} and, as expected, the two expressions agree within 1~K
in the temperature range $T_k$=1--40~K. Eq.~(4) can be employed
confidently to derive $T_k$ at temperatures lower than 40~K and for
densities smaller than $10^5 $~cm$^{-3}$. For higher densities and
temperatures, high-lying $k$-ladders (above the level $2_1$) play a
role and it is necessary to solve the statistical equilibrium
equations within a radiative transfer treatment, which is beyond the
scope of the present paper.

\section{Conclusion}\label{sec:concl}

We have presented rotation-inversion (de)excitation rate coefficients
for p- and o-NH$_3$ due to collisions with p-H$_2$($j_2 = 0$),
o-H$_2$($j_2 =1$) and H. The scattering calculations were performed
using the interaction potential described in \cite{maret09} for
NH$_3$-H$_2$ and in \cite{Guo:14} for NH$_3$-H. The lowest 17 and 34
rotational levels of o- and p-NH$_3$, respectively, were considered
and rate coefficients were computed for kinetic temperatures up to
200~K. The new rate coefficients for p-H$_2$ were found to agree
within a factor of 2 (30\% on average) compared to the older
computations of \cite{maret09}. Our calculations complete the
available sets for the ammonia molecule: specific calculations are now
available for the dominant neutral colliders of the interstellar
medium, i.e. p-H$_2$, o-H$_2$, H (this work) and He (Machin et
al. 2005).

We note that the hyperfine structure of NH$_3$ is resolved in some
astrophysical spectra. In order to model such observations, hyperfine
selective rate coefficients are necessary. 
The accurate recoupling theory (see e.g. \cite{Faure:12HFS}) should be therefore extended to the case of molecule-molecule collisions, as in \cite{lanza:14} for the HCl-H$_2$ system.
The statistical method (in which the hyperfine selective rate coefficients are taken to be proportional to the statistical weights of the final states) can be also employed
as a simple approximation. 
It should be noted, however, that the
statistical approach is known to be insufficient in some cases, as
discussed by \cite{stutzki85} and, more recently, by
\cite{Faure:12HFS}.

We note also that the present rotation-inversion rate coefficients for
the main ammonia isotopologue $^{14}$NH$_3$ can be safely employed for
the $^{15}$NH$_3$ isotopologue because the $^{14}$N/$^{15}$N isotopic
effect is negligible. In contrast, a hydrogen/deuterium substitution
has a great impact since {\it i)} the principal moment of inertia axes are
rotated (except in ND$_3$) and {\it ii)} the shift of the centre of
mass is significant. Rotational rate coefficients were computed
recently for all deuterated ammonia isotopologues (Daniel et al. 2014,
2016) so that collisional data are indeed now available for all
ammonia isotopologues.

Finally, future studies will address the excitation of higher
rotational levels, at higher temperatures, as well as the excitation
of the vibrational levels of ammonia (the lowest one opens at
950~cm$^{-1}$). This will require a full twelve-dimensional
NH$_3$-H$_2$ PES.

\section*{Acknowledgements}

This work has been supported by the Agence Nationale de la Recherche
(ANR-HYDRIDES), contract ANR-12-BS05-0011-01.

\bibliographystyle{mn2e}

\bibliography{bouhafs1105}

\bsp

\label{lastpage}

\end{document}